\documentclass[twocolumn,showpacs,preprintnumbers,amsmath,amssymb]{revtex4}

\usepackage{graphicx}
\usepackage{dcolumn}
\usepackage{bm}

\begin{document}


\title{Angular Momentum Projected Configuration Interaction with Realistic Hamiltonians}

\author{Zao-Chun Gao$^{1,2}$}%
 \email{gao1z@cmich.edu}
\author{Mihai Horoi$^1$}
\affiliation{
$^1$Department of Physics, Central Michigan University, Mount
Pleasant, Michigan 48859, USA\\
$^2$China Institute of Atomic Energy P.O. Box 275-18, Beijing 102413, China
}

\date{\today}

\begin{abstract}
The Projected Configuration Interaction (PCI) method starts from a 
collection of mean-field wave functions, and builds up correlated 
wave functions of good symmetry. It relies on the Generator 
Coordinator 
Method (GCM) techniques, but it improves the past approaches by
 a very efficient method of selecting the basis states. 
We use the same realistic Hamiltonians and model spaces 
 as the Configuration Interaction (CI) method, 
and compare the results with the 
full CI calculations in the $sd$ and $pf$ shell.  
Examples of $^{24}$Mg, $^{28}$Si, $^{48}$Cr, $^{52}$Fe and $^{56}$Ni are discussed.
\end{abstract}

\pacs{21.60.Cs,21.60.Ev,21.10.-k}
\maketitle

\section{Introduction}

The full configuration interaction (CI) method using a spherical single particle (s.p.) basis and realistic Hamiltonians 
has been very successful in describing various properties of the low-lying states in light and medium nuclei. 
The realistic Hamiltonians, such as the USD \cite{Wild84,si28}in the $sd$ shell, 
the KB3 \cite{Poves81}, FPD6 \cite{Richter91} and GXPF1 \cite{Honma02} in the $pf$ shell, 
have provided a very good base to study various nuclear structure problems microscopically.

Some $sd$ and $pf$ nuclei such as $^{28}$Si and $^{48}$Cr are well deformed. Their collective behavior 
is confirmed by the strong collective E2 transitions and the rotational behavior of the yrast
state energies, $E(I)\sim I(I+1)$. 
The mean-field description in the intrinsic frame naturally takes advantage of the spontaneous
symmetry breaking. This approach provides some physical insight, but the loss of
good angular momentum of the mean-field wave functions makes the comparison with the 
experimental data difficult.
The CI calculations in spherical basis provide the description in the laboratory frame.
The angular momentum is conserved, but the physical insight associated with the existence 
of an intrinsic state is lost. 

There is a long-lasting effort to connect the mean-field and CI techniques.
Elliott was the first to point out the advantage of a deformed intrinsic many body basis and
developed the SU(3) Shell Model \cite{Elliott58} for $sd$ nuclei.
In Elliott's model, the rotational motion was associated with strict SU(3) symmetry,
approximately realized only near $^{20}$Ne and $^{24}$Mg. This model was limited only to the $sd$ shell.

The Projected Shell Model(PSM) \cite{Hara95} can be considered as
a natural extension of the SU(3) shell model to heavier systems. 
In this model, the quadrupole force, 
that Elliott's model used, the monopole pairing and the quadrupole pairing forces 
were included in the PSM Hamiltonian. The deformed intrinsic Nilsson+BCS basis
are projected onto good angular momentum, and the PSM Hamiltonian is diagonalized 
in the space spanned by the projected states. The Nilsson model \cite{Nilsson55} has been proven 
to be very successful in describing the deformed intrinsic single particle states, and 
the quadrupole force was found to be essential for describing the rotational motion \cite{Elliott58}. 
Despite its simplicity PSM was proven to be a very efficient method in analyzing the phenomena
associated with the rotational states,
especially the high spin states, not only for axial quadrupole deformation, 
but also for the octupole \cite{Chen01} and triaxial shapes \cite{Sheikh99,Gao06}.
However, its predictive power is limited because the mulitpole-multipole plus pairing
Hamiltonian has to be tuned to a specific class of states, rather than an region of the nuclear chart.  

Besides the Projected Shell Model, another sophisticated approach based on the projection method
is MONSTER and the family of VAMPIRs \cite{Schmid04}. 
The MONSTER is similar to the PSM, but the basis is obtained by projecting  the
Hartree-Fock-Bogoliubov (HFB) vacuum and the related 2-quasiparticle
configurations onto good quantum numbers, including neutron and 
proton number, parity, and the angular momentum. 
VAMPIR, which performs the energy variation after the projection, is more sophisticated than MONSTER.
However, the particle number plus angular momentum projection 
implemented in MONSTER and VAMPIR requires at least 3-dimensional integration in the axial symmetric case.
This makes it very difficult to extend these models to non-axial cases, where the projection with
5-dimensional integration would be  needed.
MONSTER and VAMPIR can use realistic Hamiltonians.

Instead of using quasiparticle configurations, as PSM and MONSTER/VAMPIR do,
 the Quantum Monte Carlo Diagonalization (QMCD) method \cite{Honma96} takes the advantage
 of the Hartree-Fock (HF) mean-field that breaks the symmetries of Hamiltonian. 
 The mean-field is not restricted to axial symmetry. 
 QMCD starts from an appropriate initial state, 
 as in the shell-model
Monte Carlo approach \cite{Koonin97}, to select an optimal set of basis
states, and the full Hamiltonian is diagonalized
in this basis. More basis states are iteratively added until
convergence is achieved. It is very interesting that only the $M$-projection 
is applied to in the basis states, 
yet the total angular momentum seems to be fully  recovered from the diagonalization 
when the convergence limit is achieved. However, the process of selecting the basis states
is reportedly very time consuming.

The Generator Coordinate Method (GCM)\cite{HW53} is a standard method 
to describe collective states that goes beyond the mean-field. 
The angular momentum projection technique is a special case of GCM when the generator coordinates include
the Euler angles. 
In the standard GCM, the excited states are constructed in a basis of HFB vacua,
 which differ in one or several collective coordinates. 
However, in many cases, it has been found that the excitation energies provide by GCM are too high. 
The GCM wave functions may be more appropriate to describe more correlations in the ground state, rather 
than the excited states.
However, our investigations (see below) show that neither the g.s. nor the excited 
states are accurately described by this simple GCM procedure when realistic 
effective Hamiltonians and HF vacua are used.

In the present work, we developed a new method called the Projected Configuration 
Interaction (PCI), which uses GCM and PSM techniques. 
The PCI basis includes angular momentum projected states generated from a class
of constraint HF vacua.  However, in addition to these GCM-like states 
PCI includes a large number of low-lying $np-nh$ excitations built on those
states.  In PCI the deformed intrinsic states are Slater Determinants (SD), 
and the particle number projection is no longer required. 
This method can use the same realistic Hamiltonian as any CI method.
This feature makes the direct comparison between CI and PCI possible.
One advantage of this method is that it keeps the number of basis states
small, even for cases when the CI dimensions are too large for the
today's computers.

The paper is organized as follow. Section II presents the formalism used
by PCI. The efficiency of the angular momentum projection is discussed
in section III. Section IV analyzes the choice of the PCI basis and 
presents some results for the $sd$ and $pf$ nuclei. Section V is devoted
to the analysis of the quadrupole moments and E2 transitions. Conclusions
and outlook are given in section VI. 

\section{The method of the Projected Configuration Interaction (PCI)}

The model Hamiltonian used in CI calculations can be written as:

\begin{eqnarray}\label{ham}
H=\sum_i e_i c_i^\dagger c_i+\sum_{i>j,k>l}V_{ijkl}c_i^\dagger c_j^\dagger c_l c_k,
\end{eqnarray}
where, $c_i^\dagger$ and $c_i$ are creation and annihilation operators of the spherical harmonic oscillator,
 $e_i$ and $V_{ijkl}$ are one-body and two-body matrix elements that can be obtained
 from effective interaction theory, such as G-Matrix plus core polarization \cite{Morten}; 
it can be further refined using the experimental data \cite{Honma02,horoi06}.

One can introduce the deformed single particle (s.p.) basis, which can be obtained from a constraint HF solution. 
Alternatively, one can take the single particle states obtained from the following  s.p. Hamiltonian,

\begin{eqnarray}\label{hsp}
H_\text{s.p.}=h_\text{sph}-\frac23\epsilon_2\hbar\omega_0\sqrt{\frac{4\pi}5}\rho^2Y_{20}(\theta)
\nonumber\\+\epsilon_4\hbar\omega_0\sqrt{\frac{4\pi}9}\rho^2Y_{40}(\theta).
\end{eqnarray}
Here
\begin{eqnarray}\label{ho}
h_\text {sph}=\sum_i E_i c_i^\dagger c_i,
\end{eqnarray}
 is the spherical single particle part of the Hamiltonian with 
the same eigenfunctions as the spherical harmonic oscillator, but the energy of each orbit is chosen
such that the result from this Hamiltonian is near to the HF solution; $\epsilon_2$ and $\epsilon_4$ are 
the quadrupole and hexadecupole deformation parameters, 
$\rho=\sqrt{\frac{m\omega_0}{\hbar}}r$ is dimensionless, and we take \cite{Al06,gb08}
\begin{eqnarray}
\hbar \omega_0 = 45A^{1/3} - 25A^{2/3}.
\end{eqnarray}

The deformed s.p. creation operator is given by the following transformation:
\begin{eqnarray}\label{spwf}
b^\dagger_k=\sum_i W_{ki}c^\dagger_i,
\end{eqnarray}
where the matrix elements $W_{ki}=\langle b_k|c_i\rangle$ are real in our calculation. 
Inserting the reversed transformation of Eq.\,(\ref{spwf})
\begin{eqnarray}\label{revspwf}
c^\dagger_i=\sum_k W^t_{ik}b^\dagger_k
\end{eqnarray}
into the model Hamiltonian Eq. (\ref{ham}), we obtain the transformed $H$ in the deformed s.p. basis

\begin{eqnarray}\label{hdef}
H=\sum_{ij} h^{(1)}_{ij} b_i^\dagger b_j+\sum_{i>j,k>l}h^{(2)}_{ijkl}b_i^\dagger b_j^\dagger b_l b_k.
\end{eqnarray}
Here $h^{(1)}$ and $h^{(2)}$ are one-body and two-body matrix elements of $H$ in the deformed basis.
The Slater Determinant (SD) built on the deformed single particle states is given by
\begin{eqnarray}\label{sd}
|\kappa\rangle\equiv|s,\epsilon\rangle\equiv b^\dagger_{i_1}b^\dagger_{i_2}...b^\dagger_{i_n}|\rangle,
\end{eqnarray}
where $s$ refers to the Nilsson configuration, indicating the pattern of the occupied orbits, 
and $\epsilon$ is the deformation determined by $\epsilon_2$ and $\epsilon_4$.
The matrix element $\langle \kappa|H|\kappa'\rangle$, 
can be calculated using Eq.(\ref{hdef}).

Generally, $|\kappa\rangle$ doesn't have good angular momentum $I$. 
It is well known that the projection on good angular momentum would add 
correlations to the wave function.
The general form of the nuclear wave function is therefore a
linear combination of the projected SDs (PSDs),
\begin{eqnarray}\label{wv}
|\Psi^\sigma_{IM}\rangle=\sum_{K\kappa} f^{\sigma}_{IK\kappa}
P^I_{MK}|\kappa\rangle,
\end{eqnarray}
where
\begin{eqnarray}\label{pj}
\hat P^I_{MK}=\frac{2I+1}{8\pi^2}\int d\Omega D^I_{MK}(\Omega)\hat
R(\Omega)
\end{eqnarray}
is the angular momentum projection operator. $D^I_{MK}$is the
D-function, defined as
\begin{eqnarray}
D^I_{MK}(\Omega)=\langle IM|\hat R(\Omega)|IK \rangle^*,
\end{eqnarray}
$\hat R$ is the rotation operator, and $\Omega$ is the solid
angle. If one keeps the axial symmetry in the deformed
basis, $D^I_{MK}$ in Eq(\ref{pj}) reduces to 
$d^I_{MK}(\beta)=\langle IM|e^{-i\beta\hat J_y}|IK \rangle$
and the three dimensional $\Omega$ reduces to $\beta$. 

The energies
and the wave functions [given in terms of the coefficients
$f^\sigma_{IK\kappa}$ in Eq.(\ref{wv})] are obtained by solving
the following eigenvalue equation:

\begin{eqnarray}\label{eigen}
\sum_{K'\kappa'}(H_{K\kappa, K' \kappa'}^I-E^\sigma_IN_{K\kappa,
K' \kappa'}^I)f^\sigma_{IK\kappa'}=0,
\end{eqnarray}
where $H_{K\kappa, K' \kappa'}^I$ and $N_{K\kappa, K' \kappa'}^I$
are the matrix elements of the Hamiltonian and of the
norm, respectively
\begin{eqnarray}
H_{K\kappa, K' \kappa'}^I&=&\langle
\kappa|HP^I_{KK'}|\kappa'\rangle,\\ N_{K\kappa, K'
\kappa'}^I&=&\langle \kappa|P^I_{KK'}|\kappa'\rangle.
\end{eqnarray}
The matrix element of $H_{K\kappa, K' \kappa'}^I$ and $N_{K\kappa, K' \kappa'}^I$
can be expanded as
\begin{eqnarray}
&&\langle \kappa|HP^I_{KK'}|\kappa'\rangle
\nonumber\\&=&\frac{2I+1}{8\pi^2}\int d\Omega D^I_{KK'}(\Omega)\langle
\kappa|H\hat R(\Omega)|\kappa'\rangle,\\
&&\langle\kappa|P^I_{KK'}|\kappa'\rangle
\nonumber\\&=&\frac{2I+1}{8\pi^2}\int d\Omega D^I_{KK'}(\Omega)\langle
\kappa|\hat R(\Omega)|\kappa'\rangle,
\end{eqnarray}
where
\begin{eqnarray}
\langle \kappa|H\hat R(\Omega)|\kappa'\rangle=
\sum_{\kappa''} \langle \kappa|H|\kappa''\rangle\langle \kappa''|\hat R(\Omega)|\kappa'\rangle.
\end{eqnarray}
Here $\kappa''$ has the same deformation  as $\kappa$, and runs over all SDs with nonzero $\langle \kappa|H|\kappa''\rangle$. 
Therefore, the main problem is to calculate the matrix element of the rotation $\hat R(\Omega)$ between different SDs, 
$\langle \kappa|\hat R(\Omega)|\kappa'\rangle$.

$|\kappa\rangle$ and $|\kappa'\rangle$ may have different shapes.
We denote $a^\dagger,a$ the single particle operators with deformation 
$\epsilon_a$ that create the SD $|\kappa\rangle$, and $b^\dagger,b$, with deformation $\epsilon_b$,
that create  $|\kappa'\rangle$. Expressing $\langle
\kappa|\hat R(\Omega)|\kappa'\rangle$ with $a^\dagger,a$ and $b^\dagger,b$, one gets
\begin{eqnarray}\label{rme}
&&\langle \kappa|\hat R(\Omega)|\kappa'\rangle \nonumber\\&=&
\langle|a_{i_n}...a_{i_2}a_{i_1}\hat R(\Omega)b_{j_1}^\dagger b_{j_2}^\dagger...b_{j_n}^\dagger|\rangle \nonumber\\
&=&\langle|a_{i_n}...a_{i_2}a_{i_1}\nonumber\\&&\hat R(\Omega)b_{j_1}^\dagger \hat R^{-1}(\Omega)\hat R(\Omega) b_{j_2}^\dagger \hat R^{-1}(\Omega)...\hat R(\Omega)b_{j_n}^\dagger \hat R^{-1}(\Omega)|\rangle \nonumber\\
&=&\langle|a_{i_n}...a_{i_2}a_{i_1}\nonumber\\&&\left(\sum_{k_1} R_{k_1 j_1}a_{k_1}^\dagger\right)
\left(\sum_{k_2} R_{k_2 j_2}a_{k_2}^\dagger\right)...\left(\sum_{k_n} R_{k_n j_n}a_{k_n}^\dagger\right)|\rangle \nonumber\\
&=&\begin{vmatrix}
R_{i_1j_1} & R_{i_1j_2} & \dots & R_{i_1j_n}\\
R_{i_2j_1} & R_{i_2j_2} & \dots & R_{i_2j_n}\\
\hdotsfor{4}\\
R_{i_nj_1} & R_{i_nj_2} & \dots & R_{i_nj_n}
\end{vmatrix}.
\end{eqnarray}
Here
\begin{eqnarray}
&&\hat R(\Omega)b_{j}^\dagger \hat R^{-1}(\Omega)\nonumber\\
&=&\sum_l W^b_{jl}\hat R(\Omega)c_{l}^\dagger \hat R^{-1}(\Omega) \nonumber\\
&=&\sum_{lk} W^b_{jl}D^*_{kl}(\Omega)c_{k}^\dagger \nonumber\\
&=&\sum_{i}\left(\sum_{lk} W^b_{jl}D^*_{kl}(\Omega)W^a_{ik}\right)a_{i}^\dagger \nonumber\\
&=&\sum_{i} R_{ij}a_i^\dagger .
\end{eqnarray}
Note that the rotation never mixes the neutron and proton wave functions. 
Therefore, the matrix elements in Eq.(\ref{rme}) can be calculated 
for neutron and proton separately.

The choice of the PCI basis will be described in Section IV.

\section{The Efficiency of the Projection on Good Angular Momentum}
\begin{figure}
\centering
\includegraphics[width=3.0in]{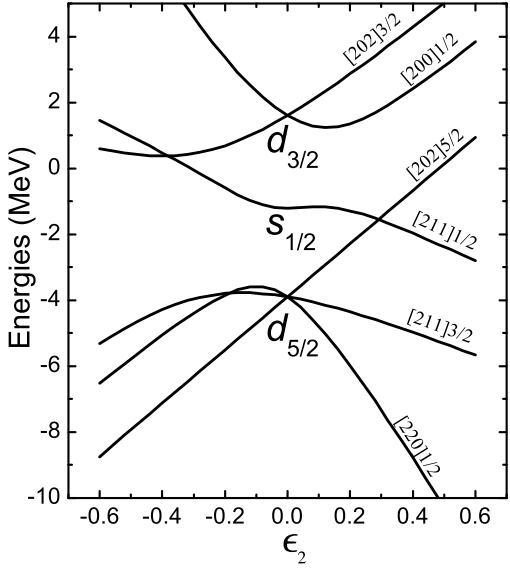}
\caption{The Nilsson levels for $sd$ shell}
\label{nilsd}
\end{figure}

Let's analyze the quality of the deformed single particle states for the USD Hamiltonian \cite{si28}. 
The $E_i$ energies (see Eq.(\ref{ho})) are properly adjusted so that the SD with the lowest energy
is close to the HF vacuum, as will be discussed in this Section.
For $sd$ shell, we use $E_{d_{5/2}}=-3.9$ MeV, $E_{s_{1/2}}=-1.2$ MeV  and $E_{d_{3/2}}=1.6$ MeV, 
and the corresponding  Nilsson levels as functions of quadrupole deformation $\epsilon_2 (\epsilon_4=0)$
are plotted in Fig.\ref{nilsd}. 

\begin{figure}
\centering
\includegraphics[width=3.0in]{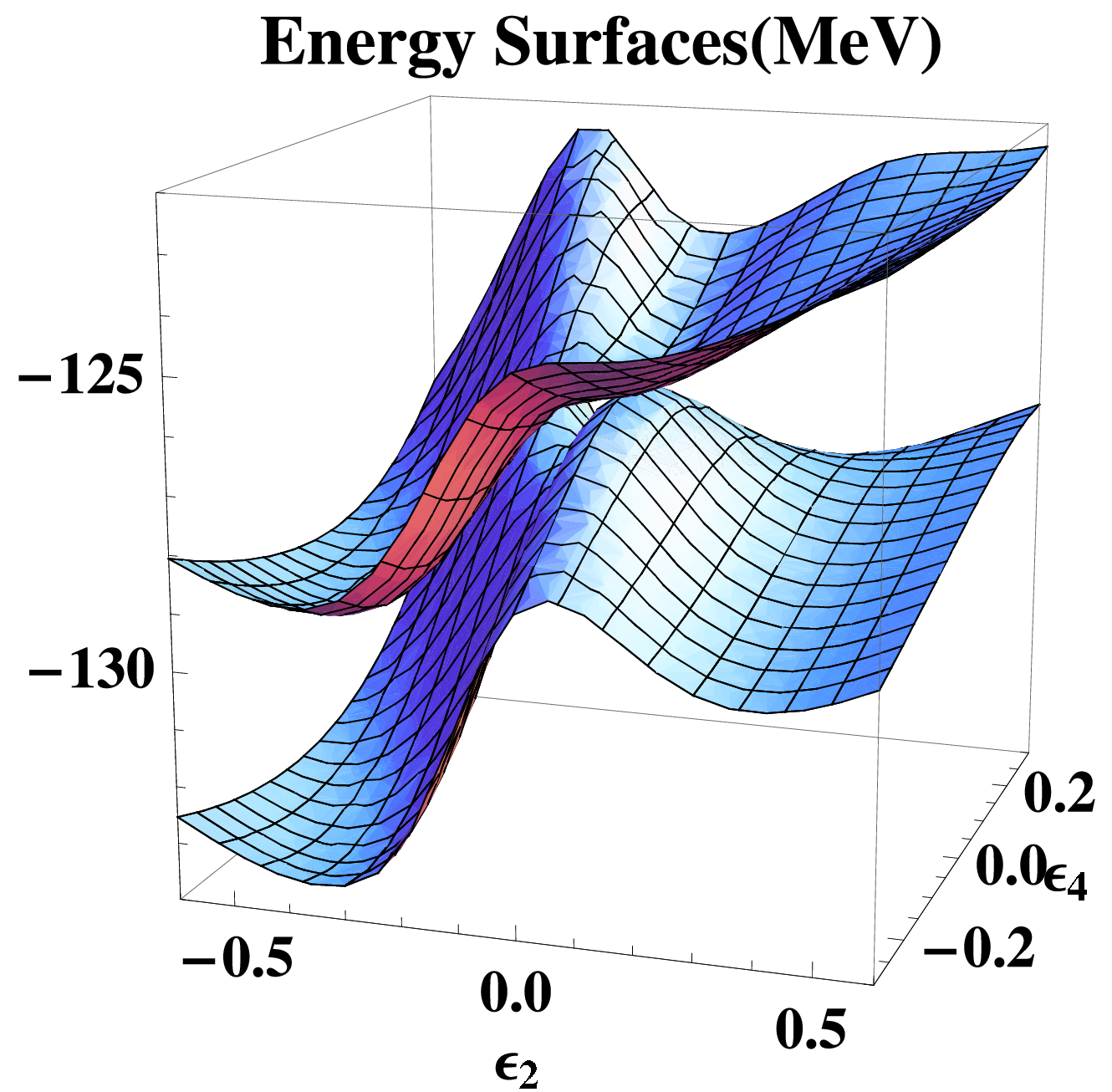}
\caption{(Color online) Energy surfaces of the USD Hamiltonian with respect to the unprojected SD($s_0$) 
(upper surface) and the projected SD($s_0$) with I=0 (lower surface),  
as functions of quadrupole $\epsilon_2$ and hexadecupole $\epsilon_4$ deformation. } 
\label{Si282s}
\end{figure}

To find the lowest value $E_\text{def}$ of $\langle \kappa|H|\kappa\rangle$, the energy surface of
\begin{eqnarray}\label{hexp}
E_\text{exp}(s,\epsilon)=\langle s, \epsilon|H|s, \epsilon\rangle
\end{eqnarray}
needs to be calculated for each configuration $s$.
Here, we use the USD Hamiltonian, and take $^{28}$Si as a case study.
The SD with all 12 particles occupying the Nilsson orbits originating from the $d_{5/2}$ spherical

orbit are used in Eq. (20). Such an SD is denoted by $s_0$ hereafter.
This configuration provides the lowest energy for a large range of deformations and is expected
to 
describe very well the mean-field minimum, i.e. the ground state (g.s.),
$\langle\kappa|H|\kappa\rangle$ at certain deformation(s). 
 The energy surface of the $s_0$ configuration
as a function of $(\epsilon_2,\epsilon_4)$ is shown as the upper surface in Fig. \ref{Si282s}. 
The minimum energy, $E_\text{def}=-129.55$ MeV at deformation $(\epsilon_2,\epsilon_4)_\text{def}=
(-0.44,-0.10)$, which
is very close to the HF energy ($E_{\text {HF}}=-129.61$ MeV) \cite{Al06}. Besides,
the calculated quadrupole moment, $-70.1$fm$^2$, is also very close the the HF value, 
$-69.6$fm$^2$ \cite{Al06}. Thus, we reached an approximated solution to the HF mean-field from a 
simple Nilsson Hamiltonian. For other deformed $sd$ nuclei the
situation is similar to $^{28}$Si, as shown in Table \ref{t1}.

The ground state of $^{28}$Si should have good angular momentum with spin $I=0$. 
We calculate the expectation value of $H$ with respect to the projected SD:
\begin{eqnarray}\label{epj}
E_\text{exp}(I,s,\epsilon)=\frac{\langle s,\epsilon|HP^I_{KK}|s,\epsilon\rangle}
{\langle s,\epsilon|P^I_{KK}|s,\epsilon\rangle},
\end{eqnarray}
where $K$ is defined by $\hat J_\text{z}|s,\epsilon\rangle=K|s,\epsilon\rangle$. 
The energy surface of $E_\text{exp}(I=0,s_0,\epsilon)$
 is shown as the lower surface in Fig. \ref{Si282s}. The minimum value $E_\text{pj}(I=0)=-133.64$ MeV 
 at $(\epsilon_2,\epsilon_4)_\text{pj}
 =(-0.40,-0.20)$, which is about 4 MeV lower than the HF energy, 
 and $2.30$ MeV above the exact ground state energy of the USD Hamiltonian at $-135.94$ MeV.
 Similar calculations for other deformed $sd$-shell nuclei are reported in Table. \ref{t1}. 
These results indicate that the restoration of the angular momentum has a significant contribution
to the ground state (g.s.) correlation energy.

\begin{table*}
\caption{\label{t1} Energies (in MeV) with USD Hamiltonian for some deformed $sd$ Nuclei. 
$E_{\text {sph}}$ is the spherical HF energy, $E_{\text {HF}}$ is the minimum of the deformed HF energy,
$E_{\text {def}}$ is the lowest energy of Eq.(\ref{hexp}) and $(\epsilon_2,\epsilon_4)_{\text {def}}$ is 
the deformation for $E_{\text {def}}$, $E_{\text {pj}}(I=0)$ is the lowest energy of Eq.(\ref{epj})
at spin $I=0$ and $(\epsilon_2,\epsilon_4)_{\text {pj}}$ is 
the deformation for $E_{\text {pj}}(I=0)$. $E_{\text {Full CI}}$ is the exact solution of the USD Hamiltonian.}
\begin{ruledtabular}
\begin{tabular}{cccccccccc}
Nucleus&$E_{\text {sph}}$\footnotemark[1]&$E_{\text {HF}}$\footnotemark[1]&$E_{\text {def}}$
&$(\epsilon_2,\epsilon_4)_{\text {def}}$&$E_{\text {pj}}(I=0)$&$(\epsilon_2,\epsilon_4)_{\text {pj}}$&$E_{\text {Full CI}}$\\
\hline
$^{20}$Ne& $-31.79$\footnotemark[2] & $-36.38$ & $-36.38$ & $(0.60,-0.04)$ & $-39.86$& $(0.56,-0.26)$  & $-40.49$ \\
$^{24}$Mg& $-68.20$ & $-80.17$ & $-79.80$ & $(0.50, 0.08)$ & $-83.04$& $(0.58, 0.04)$  & $-87.09$ \\
$^{28}$Si&$-126.03$ &$-129.61$ &$-129.55$ &$(-0.44,-0.10)$ &$-133.64$& $(-0.40,-0.20)$ &$-135.94$ \\
$^{36}$Ar&$-222.75$ &$-226.56$ &$-226.56$ &$(-0.32, 0.08)$ &$-229.78$& $(-0.20, 0.30)$ &$-230.51$ \\
\end{tabular}
\end{ruledtabular}
\footnotetext[1]{Data taken from Ref.\cite{Al06}.}
\footnotetext[2]{The $-25.46$ MeV in Ref. \cite{Al06} has been corrected as $-31.79$ MeV.}
\end{table*}

\begin{figure}
\centering
\includegraphics[width=2.5in]{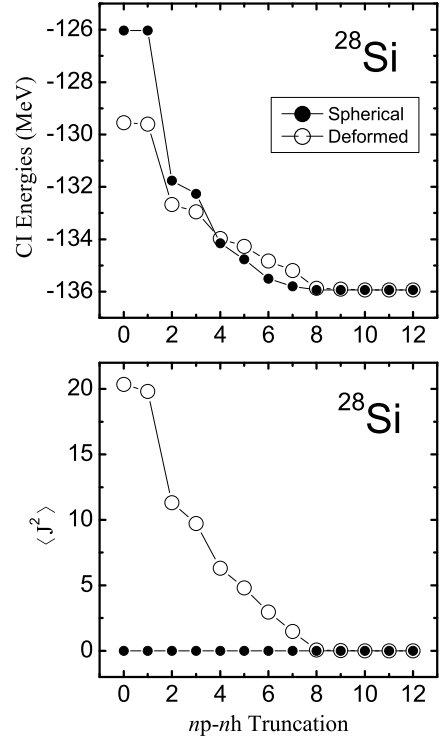}
\caption{The ground state energies (upper panel) and the corresponding
$\langle J^2\rangle$ values (lower panel) of $^{28}$Si from spherical and deformed CI
 as functions of different $n$p-$n$h truncation}
\label{Si28ph1}
\end{figure}

An even stronger argument in favor of imposing good angular momentum to the
wave functions can be found by comparing the results of CI calculations using
spherical and deformed s.p. bases, as shown in Fig. \ref{Si28ph1}.
The deformation for the deformed CI basis  is $(\epsilon_2,\epsilon_4)_\text{def}=
(-0.44,-0.10)$ as shown in Table \ref{t1}. The energies at 0p-0h are the values of $E_\text{sph}$
and $E_\text{def}$ shown in the same Table \ref{t1}. The $E_\text{def}$ is much lower than $E_\text{sph}$, 
but $\langle J^2\rangle$ for the deformed SD is as large as $20.35$, far away from $\langle J^2\rangle$=0 
of the spherical HF state. At the 2p-2h truncation level, the spherical CI energy drops very
 close to that of the deformed CI, while the latter is struggling for recovering the angular momentum, and
 $\langle J^2\rangle$ drops to $11.30$. At the 4p-4h truncation level, 
the spherical CI energy is even lower than the deformed ones, 
yet the angular momentum of the latter hasn't been completely recovered. 
The advantage of the deformed mean field is then completely lost in the CI due to the lack of good angular momentum.
At the  8p-8h truncation level the deformed CI state completely recovers its angular momentum,
 and the results from spherical CI and deformed CI become equivalent. 
Thus, we conclude that the deformed CI won't 
benefit from the deformed mean field unless the angular momentum is recovered from the outset.
The PCI method described in Section II maintains the advantages of the deformed mean field and 
of configuration mixing specific to CI, 
and could provide further improvement to the $E_{\text{pj}}(I=0)$ in Table \ref{t1}. 
 
\begin{figure}
\centering
\includegraphics[width=3.0in]{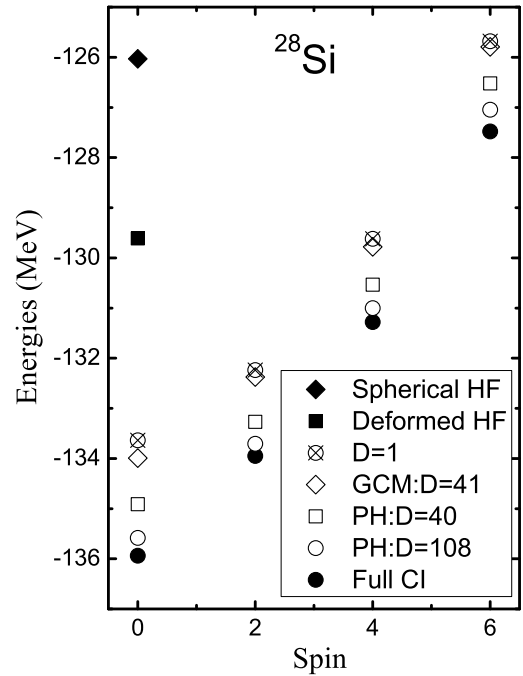}
\caption{Calculations of the ground state band in $^{28}$Si with different basis. Except for the GCM:D=41 case,
the deformation for all the other cases is $(\epsilon_2=-0.40,\epsilon_4=-0.20)$}
\label{Si28gs}
\end{figure}

To describe those non-zero spin states, one can change the spin from 0 to I in Eq. (21),
 and keep $s$ and $\epsilon$ unchanged. 
Then one gets a sequence of energies, $E_{\text{pj}}(I)$ shown as D=1 in Fig. \ref{Si28gs}.
One can see  that the values of $E_{\text{pj}}(I)$
have approximately reproduced the $I(I+1)$ feature of a rotational band in the low spin region.
The 2 MeV gap between the D=1 band and the Full CI band is expected to be reduced
by taking advantage of the CI techniques, in which the interaction between 
different intrinsic configurations
has been properly considered. 
Results of the PCI for different bases are also shown
 in Fig. \ref{Si28gs}. First, we take 41 projected SDs having the same configuration ($s_0$),
but different shapes, as in the GCM method.
 The 41 shapes considered are  taken from $\epsilon_2=-0.6$ to $0.6$ by a step of 0.03, and
  $\epsilon_4$ is obtained by minimizing the energy of Eq.(\ref{epj}) with $I=0$ at each $\epsilon_2$.
   One can see that the  calculated g.s. band is not improved. 
This can be understood by the large overlap between the PSDs in the GCM basis, 
therefore, the effective dimension of the basis is far below 41.
 Adding more PSDs with new shapes in the basis doesn't make much improvement.
Therefore, we considered up to $2p-2h$ excitations of the deformed $s_0$ 
configuration used in the D=1 case, and built a PCI basis by selecting D=40 SDs (see next section).
With this basis we solved the generalized eigenvalue
problem described in Section II. The results (PH:D=40) in Fig. \ref{Si28gs} indicate
a significant improvement over GCM (GCM:D=41).
 As the number of excited PSDs increases  up to 107, we obtain the band (labeled by PH:D=108),
 which is only about 300 keV above the band obtained with full CI.
 Therefore, we conclude that the PCI method is not only an efficient way of reducing 
 the dimension of the CI calculations, at least
in deformed nuclei, but can be also very instrumental in making the connection
between intrinsic states and laboratory wave functions.

\section{Choice of the PCI basis}

\begin{figure}
\centering
\includegraphics[width=2.8in]{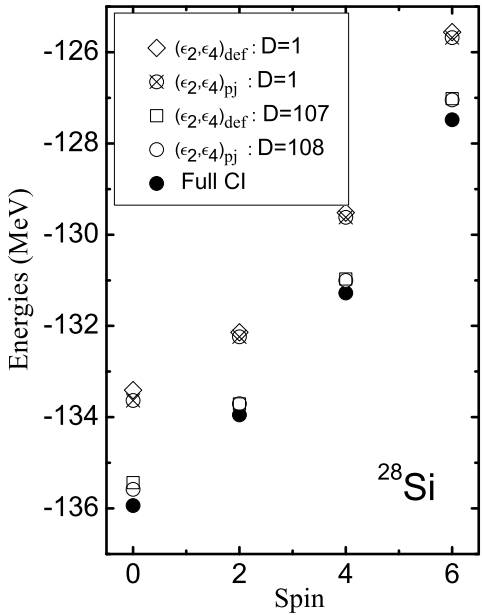}
\caption{Comparison of the calculated ground state bands in $^{28}$Si
with deformations $(\epsilon_2,\epsilon_4)_\text{pj}=(-0.40,-0.20)$ [used in Fig.\ref{Si28gs}]
and $(\epsilon_2,\epsilon_4)_\text{def}=(-0.44,-0.10)$, as shown in Table \ref{t1}}
\label{Si28def}
\end{figure}

As already mentioned, one important problem is the choice of the PCI basis.
It is  very difficult to find a set of SDs that would make an
efficient PCI basis, due to the complexity of possible structures that exist
in the nuclei, such as the spherical states, rotational states, various vibrational states, 
shape coexistence, etc. It would be helpful to be able to dentify the most 
relevant intrinsic states of a nucleus when we start a PCI calculation.
Unfortunately, there is a large number of ways of achieving that goal.
Below, we describe some of them. Assuming that we found these $N$ starting states that
we denote as $|\kappa_j,0\rangle$ ($j=1,...N$),  we further consider 
for each selected $|\kappa_j,0\rangle$ a set of relative $n$p-$n$h SDs $|\kappa_j,i\rangle$, 
and select some of them into the basis according to how much effect they can have 
on the energy of the state. Therefore, the general structure of the PCI basis is 
\begin{eqnarray}\label{basis}
\left\{\begin{matrix}
  0{\text p}-0{\text h},  & \>  n{\text p}-n{\text h} \\
|\kappa_1,0\rangle, & |\kappa_1,i_1\rangle,\cdots, \\
|\kappa_2,0\rangle, & |\kappa_2,i_1\rangle,\cdots, \\
\hdotsfor{2}\\
|\kappa_N,0\rangle, & |\kappa_N,i_1\rangle,\cdots 
\end{matrix}\right\}.
\end{eqnarray}

\begin{figure}
\centering
\includegraphics[width=3.5in]{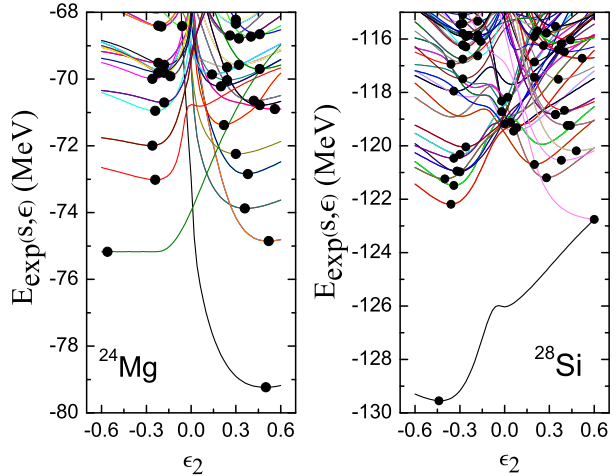}
\caption{(Color online) $E_\text{exp}(s,\epsilon)$ values
in $^{24}$Mg and $^{28}$Si, 
as functions of quadrupole deformation $\epsilon_2$. Each curve represents a Nilsson Configuration $s$, while
$\epsilon_4$ was chosen to minimize 
$E_\text{exp}(s,\epsilon)$ for each $\epsilon_2$. The absolute minimum for each $s$ is indicated by a black dot.}
\label{fig1}
\end{figure}

$N$ is somewhat arbitrary, but it can be increased until reasonable
 convergence is achieved. 
Some examples are given below.
 As a preliminary application of PCI, we consider the same set of SDs
$|\kappa_j,0\rangle$ in Eq.(\ref{basis}) for all angular momenta $I$.
The deformation of each configuration $s$ may be chosen from the minima of either
Eq. (\ref{hexp}) or Eq. (\ref{epj}) at certain $I$.
However, there seems to be not much difference between these two choices, especially 
if the studied nucleus is well deformed. 
As shown in Fig. \ref{Si28def}, the PCI results of $^{28}$Si with deformations
 $(\epsilon_2,\epsilon_4)_\text{pj}$ and $(\epsilon_2,\epsilon_4)_\text{def}$, 
 (see Table \ref{t1}) are quite close to each other. 
  Therefore, for well deformed nuclei it would be simpler to use 
Eq. (\ref{hexp}) for determining the shape of each starting configuration $s_j$.
In the present work we only consider axially symmetric shapes.
Some low-lying curves of $E_\text{exp}(s,\epsilon)$ in Eq. (\ref{hexp}) for $^{24}$Mg and $^{28}$Si 
 are plotted in Fig. \ref{fig1} as functions of $\epsilon_2$.  
$\epsilon_4$ was chosen by minimizing the $E_\text{exp}(s,\epsilon)$ for each $\epsilon_2$. 
Each configuration $s_j$ corresponds to a curve in this figure. The minima are marked by dots.

\begin{figure}
\centering
\includegraphics[height=3.5in]{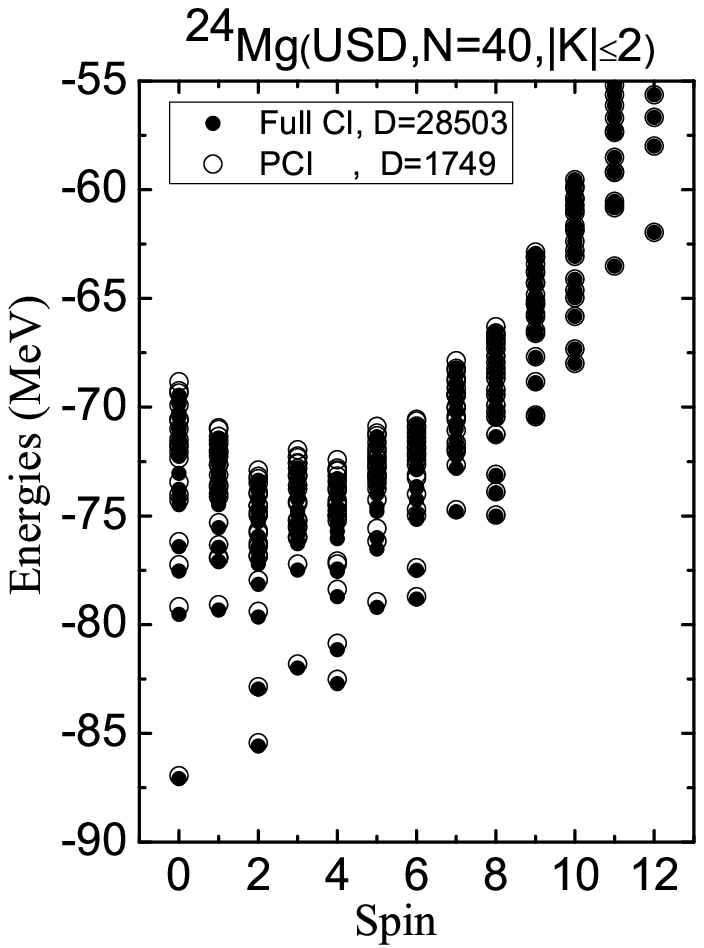}
\includegraphics[height=3.5in]{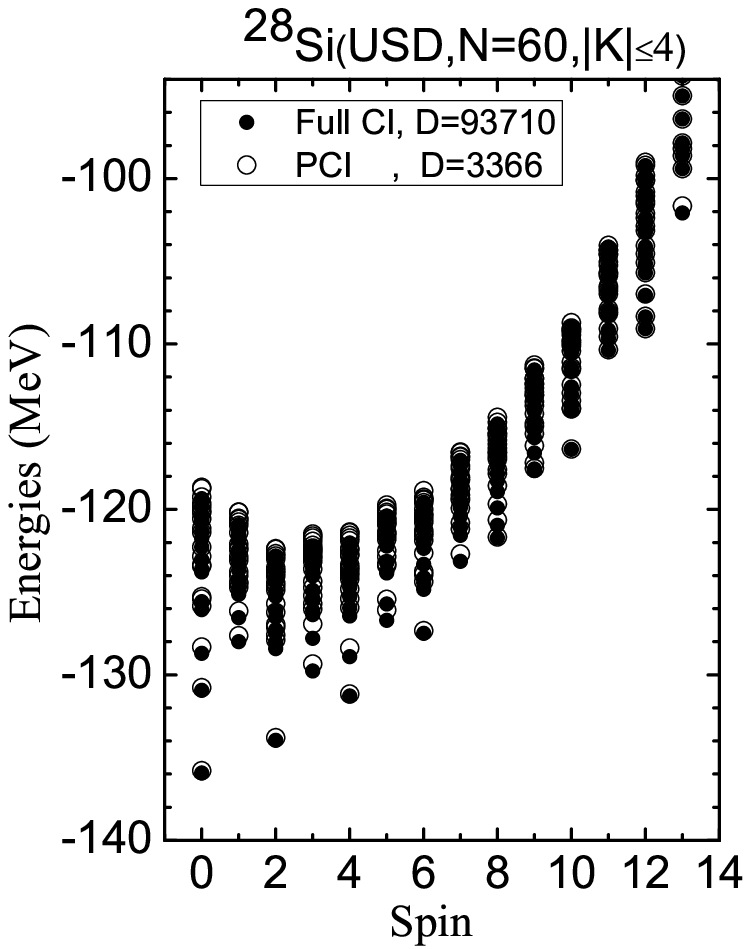}
\caption{Energies for $^{24}$Mg and $^{28}$Si calculated using PCI (open circles) and full CI(filled circles). 20
energies for each spin are presented.}
\label{fig2}
\end{figure}

In order to describe the low-lying states, 
we first choose those SDs whose energies are minima
of $E_\text{exp}(s,\epsilon)$, shown as dots in Fig. \ref{fig1}, 
and take them as starting $|\kappa_j,0\rangle$ states in Eq. (\ref{basis}).
Secondly, we try to include CI-like correlations. Each $|\kappa_j,0\rangle$ 
may have correlations with its 1p-1h and 2p-2h SDs, $|\kappa_j,i\rangle$,
through the non-diagonal matrix elements of $H$. For each $j$ we include those SDs 
that satisfy the following criterion:
\begin{eqnarray}
\Delta E=\frac12(E_0-E_i+\sqrt{(E_0-E_i)^2+4|V|^2})\geq E_\text{cut},
\end{eqnarray}
where $E_0=\langle\kappa,0|H|\kappa,0\rangle$, $E_i=\langle\kappa,i|H|\kappa,i\rangle$ and
$V=\langle\kappa,0|H|\kappa,i\rangle$. (We skipped the subscript $j$ to keep notations short.)

\begin{figure}
\centering
\includegraphics[height=3.5in]{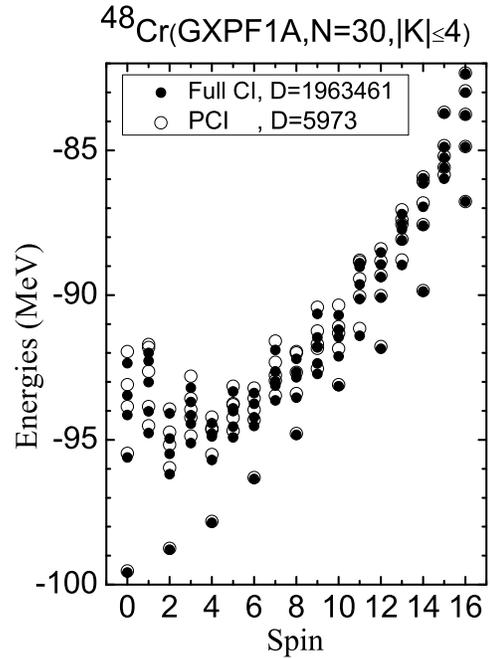}
\caption{The lowest 5 energies at each spin for $^{48}$Cr calculated using PCI (open circles) and full CI (filled circles).}
\label{Cr48}
\end{figure}

\begin{figure}
\centering
\includegraphics[width=3.5in]{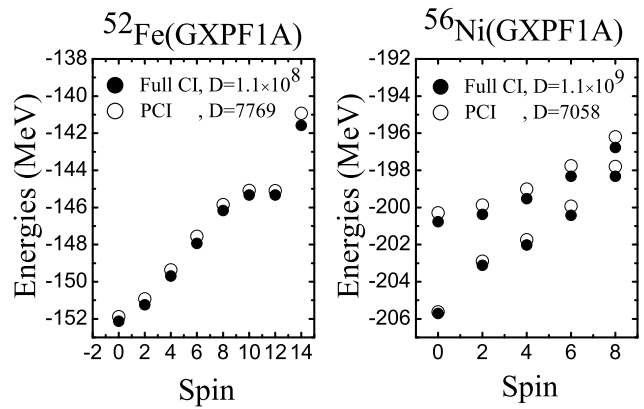}
\caption{Calculated yrast band energies using PCI (open circles) and full CI (filled circles) for $^{52}$Fe and $^{56}$Ni.
The deformed band in $^{56}$Ni starting around 5 MeV is also shown (see text for details).}
\label{FeNi}
\end{figure}

In this way we constructed the basis for the PCI calculations of $^{24}$Mg 
and $^{28}$Si in the $sd$ shell. 
For the $^{24}$Mg, we included $N=40$ lowest energy
$|\kappa_j,0\rangle$ SDs with $|K|\le 2$, and
1749 SDs are selected for the basis when $E_\text{cut}=10$ keV.
For the $^{28}$Si case, $N=60$, $|K|\le 4$ and $E_\text{cut}=10$ keV selects 3366 SDs.
The choice of $|K|\le 2$ seems to be arbitrary, and at this point it is related to the
number of states one wants to include in the basis. One should emphasize that the PCI method 
combines the advantage of using a  set of s.p. bases of different deformations with that of
the CI $np-nh$ configuration mixing. In m-scheme CI, for example, one can get the states for all
spins using the lowest M value (0 for even-even cases), although using higher M values could
be somewhat more efficient for higher spins. In PCI one has to weigh in the advantages of using
a larger number of np-nh excitations with those of using larger K values for larger spins.
Using our choices for the K values the lowest 20 PCI energies of each 
spin are compared in Fig. \ref{fig2} with the full CI results. 
Almost all low-lying energies for  each spin reproduce the full CI results very well. 
We conclude that the selected PCI basis has already carried almost the 
whole information of the low-lying states,
yet we chose only a small number of SDs as compared with those used in the full $m$-scheme CI.

The QMCD \cite{Honma96} method also uses a relatively small number of selected SDs. 
For a basis of 800 SDs  QMCD found a g.s. energy of $-89.91$ MeV for $^{24}$Mg.
In the present calculation, after the orthogonalization of the 524 $K=0$ SDs, 453 SDs are used 
to calculate the I=0 energies (see Table \ref{dim}).
The resulting PCI g.s. energy is $-89.94$ MeV. 
Furthermore, the QMCD paper only reports the 6 lowest states with spin up to $I=4$, 
while our PCI calculation with
the selected 1749 SDs can provide tens of excited states with spin running from 0 to 12 that compare
well with the full CI results.
Our approach seems to largely extended the range of nuclear states that can be 
calculated using a relatively small basis.

The results for $^{28}$Si are similar to those of $^{24}$Mg. 
Comparing with the $N=1$ calculation,
that has been discussed in Section III, the dimension is much larger 
because one needs to 
describe a large number of low-lying states of the same spin. 
For $I=0$  the number of $K=0$ $|\kappa_j,0\rangle$ SDs is $N=15$, 
and a total of 747 independant SDs are selected from the 886 $K=0$ SDs.
The first 20 $0^+$ states, up to about 18 MeV excitation energy, 
are reasonably close to the full CI results. 
For $I\neq 0$ the situation is very much the same.

The $pf$ calculations show similar trends. 
For $^{48}$Cr, the CI $m$-scheme dimension is 1963461. Full CI calculations
can reproduce the rotational nature of the yrast band and its backbending \cite{Cr95}. 
With PCI we choose $N=30$, $|K|\leq 4$, and for $E_\text{cut}=0.2$ keV 5973 SDs were selected. 
The lowest 5 energies for each spin are compared with the full CI results in Fig. \ref{Cr48}.
Again, we reached a satisfactory agreement between PCI and full CI.

$^{52}$Fe and $^{56}$Ni are not good rotors, and the present choice of the 
basis is not so efficient as for $^{48}$Cr. However, one can still select a PCI basis that can 
provide a reasonable description of the yrast energies when compared 
with the full CI calculations,
as shown in Fig.\ref{FeNi}. The number of deformed SDs used in these calculations is about 7000, 
which is just a small fraction of the full CI m-scheme dimension for $^{52}$Fe, 
and for $^{56}$Ni (see Table III). 
However, the choice of the PCI basis for $^{52}$Fe and $^{56}$Ni needs to be improved. 
Details will be presented in the forthcoming paper.

For $^{56}$Ni a rotational band has been identified in experiment \cite{Ni99} that starts
at around 5 MeV excitation energy. The band was successfully described by full CI
calculations in the $pf$-shell using the GXPF1A interaction\cite{horoi06}.
It was also shown \cite{horoi06} that this band has a 4p-4h character, but in order to 
get a good description of the band 10p-10h excitations from $(f_{7/2})^{16}$ configuration
are necessary.
In the present PCI calculation this rotational band has been identified (see Fig. 9) in the calculated 
excited states by inspecting the quadrupole moments and the $B(E2)$ transitions (See Table \ref{ni56em}).
The results seem to be in reasonable good agreement with those of the full CI calculations.
The details about the calculation of the quadrupole moments and the $B(E2)$ transitions are given in the
next section.

\begin{table}
\caption{\label{ni56em} Quadrupole moments (in $e$fm$^2$) and $B(E2,I\rightarrow I-2)$ (in $e^2$fm$^4$) 
values of the states in the deformed band of $^{56}$Ni calculated with PCI and CI. 
The energies for $I\neq0$ are relative to the $I=0$ bandhead.
The CI values are taken from Ref. \cite{horoi06}}
\begin{ruledtabular}
\begin{tabular}{ccccccc}
Spin& \multicolumn{2}{c}{Energy (MeV)}&\multicolumn{2}{c}{$Q(I)$}&\multicolumn{2}{c}{$B(E2)$}\\
\cline{2-3}\cline{4-5}\cline{6-7}
 &PCI&CI&PCI&CI&PCI&CI\\
\hline
0&-200.295&-200.762& 0 & 0 & &\\
2&0.425&0.395&  -48.9&-41.6 & 567.1&413.2\\
4&1.312&1.241&  -44.0&-55.2 & 646.0&598.0\\
6&2.541&2.448&  -70.5&-56.2 & 708.0&609.3\\
8&4.103&3.991&  -74.5&-47.2 & 938.6&558.4\\
%
%
\end{tabular}
\end{ruledtabular}
\end{table}

The number of SDs selected for the PCI calculation are shown in the corresponding figures
and compared with the full CI m-scheme dimensions.
However, the efficiency of the PCI truncation can also be assessed by comparing the PCI dimensions 
with the coupled-I CI dimensions.
In PCI the states used for the diagonalization of each spin in Eq. (12) are obtained from the selected PSDs 
after the latter are orthogonalized and the redundant states are filtered out.
The remaining states form the PCI basis for each spin. These dimensions are compared in Table \ref{dim}  with the 
full coupled-I CI dimensions. One can see that PCI can provide reasonable good results using much smaller
dimensions than the corresponding m-scheme and/or coupled-I CI calculations.

\begin{table}
\caption{\label{dim} PCI Dimensions for several spins compared with those of full CI for the nuclei described in this Section.}
\begin{ruledtabular}
\begin{tabular}{cccccc}
Nucleus&&$I=0$&$I=2$&$I=4$&$I=6$\\
\hline
$^{24}$Mg&SM& 1161 & 4518 & 4734 & 2799\\
         & PCI  & 453  & 1569 & 1579 & 1492\\
$^{28}$Si&SM& 3372 &13562 & 15089 & 9900\\
         & PCI  &  747 & 2266 &  2919 & 2711\\
$^{48}$Cr&SM& 41355&182421& 246979& 226259\\
         & PCI  &  1671& 3888 & 5618  & 5594\\
$^{52}$Fe&SM&$1.8\times 10^6$&$8.0\times 10^6$&$1.2\times 10^7$&$1.2\times 10^7$\\
         & PCI  &4288& 6422 & 6811  & 6806\\
$^{56}$Ni&SM& $1.5\times 10^7$ & $7.1\times 10^7$ & $1.1\times 10^8 $&$1.1\times 10^8$\\
         & PCI  & 2373 & 3968 & 5452  & 6231
\end{tabular}
\end{ruledtabular}
\end{table}

Although the PCI calculations could have much smaller dimensions than those of the corresponding full CI calculations,
the $H$ and $N$ matrix in Eq. (12) are dense. To get an idea of the workload of a PCI calculation, the
time necessary to calculate the $H$ and $N$ matrices for $^{56}$Ni at I=0, 2, 4, 6, 8 takes around 8 
hours using one processor, and the calculation of 5 eigenvalues for each spin takes about 2 hour, when
the generalized Lanczos method is used. As a comparison, the full CI calculations of the same states using the
modern coupled-I code NuShellX \cite{nushellx} could take several weeks, when one processor is used. One should also observe 
that the PCI computational load is shifted towards the calculation of the matrix elements, which can be more
efficiently parallelized than any large matrix eigenvalue solver.

\section{Quadrupole Moments and BE2 Transitions}

The quadrupole moments and $B(E2)$ transitions are given by
\begin{eqnarray}\label{q}
Q(I)&=&\sqrt{\frac{16\pi}5}\langle \Psi_{IM=I}|\hat Q_{20}|\Psi_{IM=I}\rangle \nonumber\\
&=&\sqrt{\frac{16\pi}5}\langle II,20|II\rangle\langle \Psi_I||\hat Q_2||\Psi_I\rangle
\end{eqnarray}
\begin{eqnarray}\label{be2}
B(E2;I_\text{i}\xrightarrow{} {I_\text{f}})&=&\frac{2I_f+1}{2I_i+1}|\langle \Psi_{I_f}||\hat Q_2||\Psi_{I_i}\rangle|^2 \nonumber\\
&=&|\langle \Psi_{I_i}||\hat Q_2||\Psi_{I_f}\rangle|^2
\end{eqnarray}

where 
\begin{eqnarray}
\hat Q_{\lambda \mu}=e_{n(p)} r^2Y_{\lambda \mu}(\theta,\phi),
\end{eqnarray}
with effective charges taken as $e_n=0.5e$, $e_p=1.5e$. The reduced matrix element in Eq.(\ref{q}) and (\ref{be2})
 can be expressed in terms of the PCI wave functions
\begin{eqnarray}
&&\langle \Psi_{I_f}||\hat Q_2||\Psi_{I_i}\rangle\nonumber\\&=&\sum_{K\kappa,K'\kappa'}f_{I_iK\kappa}f_{I_fK'\kappa'}
R_{I_fK'\kappa',I_iK\kappa},
\end{eqnarray}
where
\begin{eqnarray}
&&R_{I_fK'\kappa',I_iK\kappa}\nonumber\\
&=&\sum_\nu \langle I_iK'-\nu,\lambda\nu|I_fK'\rangle\langle
\Phi_{\kappa'}|\hat Q_{\lambda\nu}P^{I_i}_{K'-\nu,K}|\Phi_{\kappa}\rangle.
\end{eqnarray}
Explicit expression for the 
$\langle \Phi_{\kappa'}|\hat Q_{\lambda\nu}P^{I_i}_{K'-\nu,K}|\Phi_{\kappa}\rangle$ 
can be found in Ref. \cite{Hara95}. 
Here $\langle I_iK'-\nu,\lambda\nu|I_fK'\rangle$ is a Clebsh-Gordon coefficient.

\begin{figure}
\centering
\includegraphics[width=3.5in]{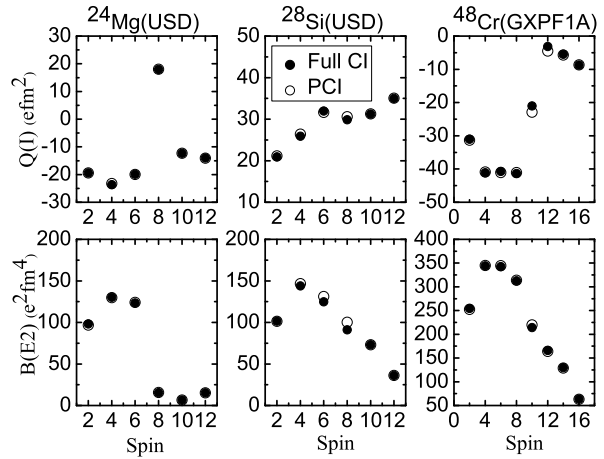}
\caption{Calculated quadrupole moments (upper panel) and $B(E2)$ (lower panel)
 from PCI (open circles) and full CI (filled circles)for $^{24}$Mg, $^{28}$Si and $^{48}$Cr.}
\label{E2}
\end{figure}

Using the PCI wave functions for the states shown in Fig.\ref{fig2} and Fig.\ref{Cr48}, 
we calculated the $Q(I)$ and the $B(E2)$ values for the yrast band in $^{24}$Mg,
$^{28}$Si and $^{48}$Cr. The results  are shown in Fig.\ref{E2}.
The very good agreement between PCI and full CI indicates that the PCI wave functions 
are almost equivalent to the exact ones.

\section{Conclusions and Outlook}

In this paper we investigate the adequacy of using a deformed s.p. basis for CI calculations.
We show that even if the starting energy of the deformed mean-field is lower
than the spherical one, the series of $np-nh$ truncations in the deformed basis
exhibits slower convergence than that of the spherical basis, due to the loss
of rotational invariance that is not recovering fast enough.

Therefore, we propose a new strategy called Projected Configuration Interaction (PCI) 
that marries features of GCM, by projecting on good angular momentum SDs built with deformed
s.p. orbitals, with CI techniques, by mixing a large number of appropriate 
deformed configurations and their $np-nh$ excitations. This ansatz seems to be very successful
in describing not only the yrast band, but also a large class of low-lying non-yrast states.
This method seems to work extremely well for deformed nuclei, and it needs further 
improvements to the choice of the PCI basis for spherical nuclei.

For the $sd$ and $pf$-shell nuclei that we studied we show that the quadrupole moments and the 
BE(2) transition probabilities are very well reproduced, indicating that not only 
the energies, but also the wave functions are accurately described. This suggests that
this method could be helpful in finding relations between the intrinsic states, which offer
more physics insight, and the laboratory wave functions provided by CI.

One of the simplifications that makes these calculations with tens of thousands of
projected states possible is the approximation of the deformed mean-field with an 
axially symmetric rotor defined by its quadrupole and hexadecupole deformations.
This approach seems to describe very well the HF mean field for the $sd$ and $pf$-shell
nuclei. We plan to extend our method to include octupole deformation, and to consider 
triaxial shapes.

\begin{acknowledgments}

The authors acknowledge support from the DOE Grant No. DE-FC02-07ER41457.
M.H. acknowledges support from NSF Grant No. PHY-0758099.
Z.G. acknowledges the NSF of China Contract Nos. 10775182, 10435010 and 10475115.
\end{acknowledgments}


\newpage
\begin{thebibliography}{10}
\bibitem{Wild84} B. H. Wildenthal, Prog. Part. Nucl. Phys. {\bf 11}, 5 (1984).
\bibitem{si28} B.A. Brown and B.H. Wildenthal, Ann. Rev. Nucl. Part. Sci.
{\bf 38}, 29 (1988).
\bibitem{Poves81} A. Poves, and A. P. Zuker, 1981a, Phys. Rep. {\bf 71}, 141.
\bibitem{Richter91} W. A. Richter, M. J. van der Merwe, R. E. Julies, and B. A.
Brown, 1991, Nucl. Phys. A {\bf 523}, 325.
\bibitem{Honma02} M. Honma, T. Otsuka, B. A. Brown, and T. Mizusaki,
Phys. Rev. C {\bf 65}, 061301(R)(2002).
\bibitem{Elliott58} J. P. Elliott,  Proc. R. Soc. London, Ser. A {\bf 245}, 128,562(1958).
\bibitem{Hara95} K. Hara and Y. Sun, Int. J. Mod. Phys. E{\bf 4},637(1995).
\bibitem{Nilsson55}S. G. Nilsson. Dan. Mat. Fys. Medd {\bf 29} nr.16(1955)
\bibitem{Chen01}Y. S. Chen and Z. C. Gao, Phys. Rev. C {\bf 63} 014314(2000).
\bibitem{Sheikh99}J.A. Sheikh and K. Hara, Phys. Rev. Lett. {\bf 82}, 3968(1999).
\bibitem{Gao06} Z. C. Gao, Y. S. Chen, and Y. Sun, Phys. Lett. B{\bf 634} 195(2006)
\bibitem{Schmid04}K.W. Schmid, Prog. Part. Nucl. Phys. {\bf 52}, 565(2004) 
\bibitem{Honma96} M. Honma, T. Mizusaki, and T. Otsuka, Phys. Rev. Lett.
{\bf 77}, 3315 (1996).
\bibitem{Koonin97} S. E. Koonin, D. J. Dean, and K. Langanke,  Phys. Rep.
{\bf 278}, 2(1997).
\bibitem{HW53}D. L. Hill and J. A. Wheeler, 1953, Phys. Rev. 89, 1102.
\bibitem{Morten} M. Hjorth-Jensen, T. T. S. Kuo and E. Osnes, Phys. Rep. {\bf 261}, 125 (1995).
\bibitem{horoi06} M. Horoi, B. A. Brown, T. Otsuka, M. Honma, and T. Mizusaki,
Phys. Rev. C{\bf 73} 061305(R)(2006).
\bibitem{Al06} Y. Alhassid, G. F. Bertsch, L. Fang and B. Sabbey, Phys. Rev. C{\bf 74} 034301(2006)
\bibitem{gb08} R. Rodríguez-Guzm$\acute{a}$n, Y. Alhassid and G.F. Bertsch, Phys. Rev. {\bf 77}, 064308 (2008). 
\bibitem{Cr95} E. Caurier et al., Phys. Rev. Lett. 75, 2466 (1995).
\bibitem{Ni99} D. Rudolph et al., Phys. Rev. Lett. 82, 3763 (1999).
\bibitem{nushellx} W.D.M. Rae, NuShellX code 2008, http://knollhouse.org/NuShellX.aspx
\end{thebibliography}
\end{document}